\documentclass[a4paper,11pt]{article}
\usepackage{pos}

\usepackage{epstopdf}
\def\pt{$p_t$}

\def\vtwopt{$v_2(p_t)$}
\def\yt{$y_t$}
\def\nch{$n_{ch}$}
\def\mt{$m_t$}

\def\k0s{$\text{K}^0_\text{S}$}
\def\mmpt{$\bar p_t$}
\def\mmpt{$\bar p_t$}

\def\pp{\mbox{$p$-$p$}}
\def\ppb{\mbox{$p$-Pb}}
\def\auau{\mbox{Au-Au}}
\def\pbpb{\mbox{Pb-Pb}}
\def\bea{\begin{eqnarray}}
\def\eea{\end{eqnarray}}
\def\nn{\mbox{N-N}}
\def\aa{\mbox{A-A}}

\title{Few-gluon interactions in high-energy  collisions}

\author*[a]{Thomas A. Trainor}

\affiliation[a]{University of Washington,\\
 Seattle, USA}

\emailAdd{ttrainor99@gmail.com}

\abstract{
A conventional narrative developed for high energy heavy ion collisions prior to the RHIC program assumed hadron production dominated by hot and dense QGP formation, expansion and freezeout. Proposed diagnostics included detection of elliptic and radial flows via correlation and spectrum analysis and jet quenching in a dense medium via rescaled spectrum ratios. The validity of that narrative has come into question in recent years with detection of  similar features in data from small collision systems. In response, several recent observations hint at a strongly modified narrative. Study of \ppb\ spectra has revealed that the Glauber Monte Carlo estimation of \mbox{A-B} collision geometry is quite misleading. Collision geometry should instead be inferred by inverting ensemble-mean \mmpt\ data. A method has been developed to transform $v_2(p_t)$ data to {\em quadrupole spectra} $\bar \rho_2(p_t)$ that provide new insights into the true correlation mechanism. Accurate \pt-integrated $v_2$ vs \nch\ measurements for \pp\ collisions reveal a {\em cubic} trend suggesting a {\em three-gluon} interaction as the true mechanism. The \pp\ $v_2(n_{ch})$ trend has been generalized to describe all \mbox{A-B} collision systems in common in terms of an {\em extensive} measure $V_2^2 \equiv \bar \rho_0^2 v_2^2$ (number of correlated pairs). A similar program has been applied to jet-related contributions to spectra and angular correlations with a comparable result, a {\em quadratic} trend on \nch\  as expected for {\em two-gluon} interactions. Those new results suggest that the relevant data features actually represent {\em multipole} (color dipole and quadrupole) radiation from few-gluon interactions within a low-density system.
}

\FullConference{The 54th International Symposium on Multiparticle Dynamics (ISMD2026)\\
17-21 August, 2026\\
National Central University, Chungli, Taiwan\\}

\begin{document}
\maketitle

\section{Introduction}

This article summarizes a presentation on few-gluon interactions within A-B collisions at the RHIC and LHC that result in production of correlated structures including dijets and an azimuth quadrupole conventionally associated with  \vtwopt. Topics include two contexts for the azimuth quadrupole, inference of quadrupole spectra from \pbpb\ \vtwopt\ data, a two-component \pt\ spectrum model, estimation of \aa\ collision geometry, model fits to 2D angular autocorrelations and A-B trends for color multipole radiation. Results reported here are based on material in Refs.~\cite{tomnewquad,tompbpb,jetspid}.


\section{Azimuth quadrupole phenomenon vs $v_2(p_t)$ data -- two contexts} \label{poskvol}

The correlation data manifestation within high-energy collisions near midrapidity conventionally identified with ``elliptic flow'' is, without reference to a physical origin, an {\em azimuth cylindrical quadrupole}. Nucleon flow observed at the Bevalac was correctly identified via the sphericity tensor~\cite{sphere}. An extrapolation to collider energies associated a {\em component} of the sphericity tensor with the symbol $v_2$ assuming that nucleon flow at the Bevalac might be extrapolated to parton flow (QGP) at collider energies~\cite{olli}.
The quantity $v_2(p_t)$ was defined in widely-cited Ref.~\cite{poskvol} as a ratio
\bea \label{v2av} 
v_2(p_t) &=&  \frac{\int_0^{2\pi} d\phi_r  \cos(2\phi_r) \bar \rho(p_t,\phi_r)}{\int_0^{2\pi} d\phi_r  \bar \rho(p_t,\phi_r)},
\\ \nonumber
&\rightarrow & \langle  \cos(2\phi_r) \rangle(p_t)
\eea
assuming a {\em common source} for almost all hadrons represented by the particle density $\bar \rho(p_t,\phi_r)$. That assumption leads to the second line as a consequence. The assumed common source is the sought-after quark-gluon plasma or QGP, a desired result thus {\em built into the analysis infrastructure}. It is assumed that the collision system includes a {\em reaction plane} from which $v_2(p_t)$ may be inferred.

Within the real high-energy physics (HEP) world the ratio $v_2(p_t)$ should be reexpressed as 
\bea \label{v2struct}
v_2(p_t) \hspace{-.05in} &=& \hspace{-.05in} \frac{V_2\{\rm 2D\}(p_t) + \text{jet contribution}~ (nonflow)}{(N_{part}/2)\bar \rho_{sNN}(p_t) + N_{bin} \bar \rho_{hNN}(p_t)},
\eea
where the numerator (at least the $V_2$ component) is a Fourier coefficient representing angular correlations and the denominator is the single-particle (SP) \pt\ spectrum $\bar \rho_{0}(p_t)$ (in a TCM format) with $v_2(p_t) \neq \langle  \cos(2\phi_r) \rangle(p_t)$. As will become apparent below $v_2(p_t)$, as the square root of a pair ratio,  depends on {\em three distinct particle sources} and has generated a great deal of confusion.

\section{Quadrupole source radiation field}

The numerator of $v_2(p_t)$ includes $V_2\{\rm 2D\}(p_t)$ that may be accurately isolated via 2D model fits to {\em autocorrelation functions}, pair distributions on the difference-variable space $(\eta_\Delta,\phi_\Delta)$, where $x_\Delta \equiv x_i - x_j$ for particles $i$ and $j$. As a Fourier coefficient it may be expressed as the first line of
\bea \label{combfac}
V_{2i}(y_t,\Delta y_{t0},\Delta y_{t2})
&\equiv&\frac{1}{2\pi} \int_{-\pi}^{\pi} d\phi_r \ \cos(2 \phi_r)  \bar \rho_{2i}(y_t,\phi_r)
\\ \nonumber
 &\approx&\bar \rho_{0i}(y_t) v_{2i}(y_t)  ~~\text{measured data, from Eq.~(\ref{v2struct})}
\\ \nonumber  
&\approx& 
 \hspace{-.04in}p'_t\, \frac{ \Delta y_{t2}}{2T_2} \, \bar \rho_{2i}(y_{t},\Delta y_{t0}) ~~~\text{important new factor $p_t'$ discovered}.
\eea
The second line introduces an inverse problem. The Fourier integral is approximately equal to measured product $\bar \rho_{0i} v_{2i}(p_t)$. The unknown is density $\bar \rho_{2i}(y_t,\phi_t)$ describing the radiation field of a quadrupole source. The inverse problem may be solved by {\em forward integration} utilizing a {\em conjectured source model} -- in this case a Boltzmann exponential on transverse mass $m_{ti}'$ where primed quantities are in the boost frame. The boost model is a combination of monopole $\Delta y_{t0}$ (``radial flow'') and quadrupole  $\Delta y_{t2}$ (``elliptic flow'') boost components. The third line above, revealing new factor $p_t'$, is the solution that can be rearranged to define a {\em quadrupole spectrum}
\bea \label{rho2i}
 \bar \rho_{2i}(y_{ti},\Delta y_{t0}) &\propto& \frac{1}{p_t'}  \rho_{0i}(y_{ti}) v_{2i}(y_{ti}).
\eea
Quantity $p_t'$ is determined exactly if a monopole boost $\Delta y_{t0}$ value is obtained from \vtwopt\ data. A quadrupole spectrum may thus be inferred with few percent accuracy from measured data as established in Ref.~\cite{quadspec}. The next section reports recent analysis of 2.76 TeV \pbpb\ $v_{2i}(p_t)$ data~\cite{tomnewquad}.


\section{2.76 TeV Pb-Pb $v_2(p_t)$ data}

\begin{figure}[h]
     \includegraphics[width=2.9in]{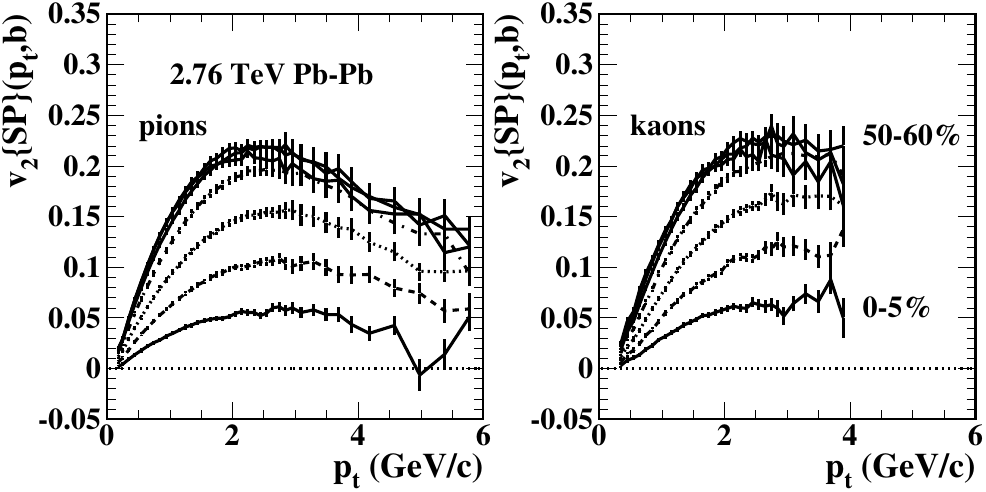}
\put(-127,90) {\bf (a)}
\put(-22,90) {\bf (b)}
    \includegraphics[width=2.9in]{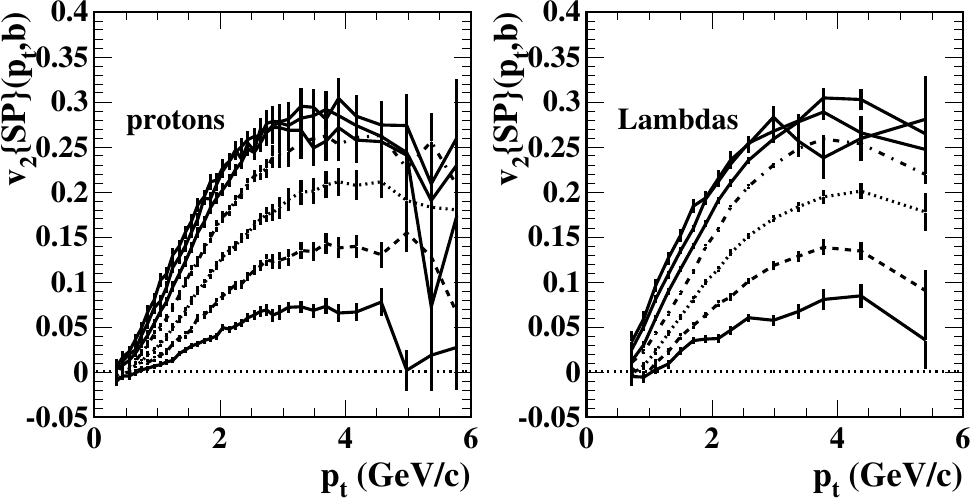}
\put(-127,90) {\bf (c)}
\put(-22,90) {\bf (d)}\\
     \includegraphics[width=2.9in]{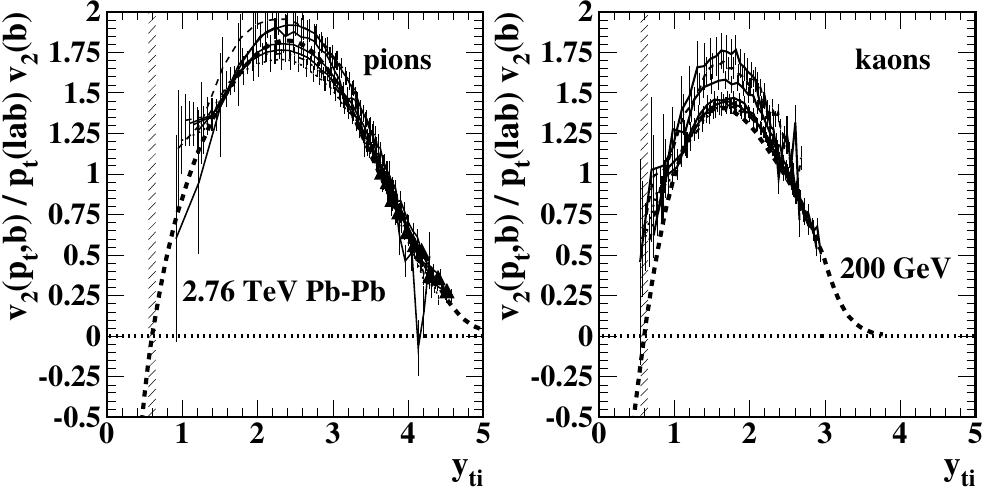}
 \put(-125,20) {\bf (a)}
\put(-22,20) {\bf (b)}
    \includegraphics[width=2.9in]{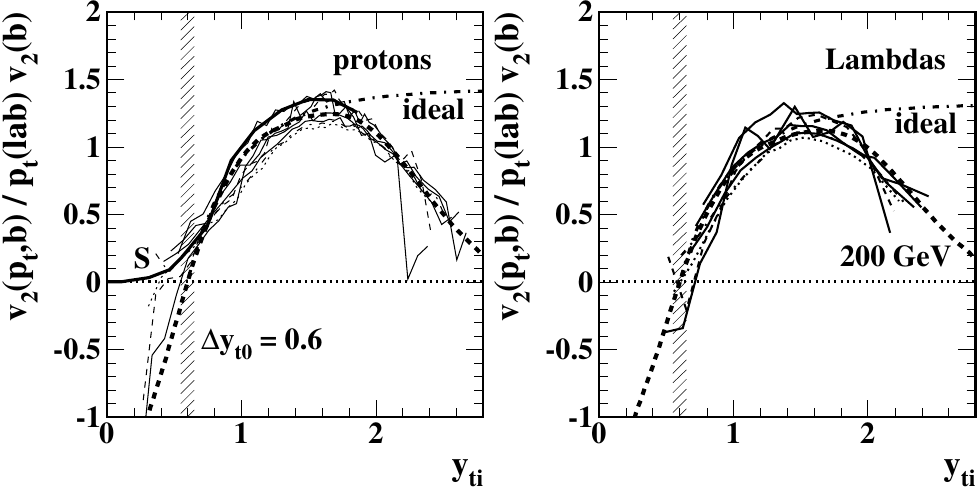}
\put(-125,20) {\bf (c)}
\put(-21,20) {\bf (d)}
\caption{\label{alice1}
First row: $v_2(p_t,n_{ch})$ data for four hadron species from seven centrality classes of 2.76 TeV \pbpb\ collisions~\cite{alicev2ptb}. Bars represent statistical and systematic errors combined quadratically. Line styles proceed from most-central data as solid, dashed, dotted, dash-dotted with solid thereafter.
Second row: Data transformed as noted above and shifted on \yt\ to a common source boost $\Delta y_{t0} = 0.6$. 
Solid curve S in panel (c) is a viscous-hydro prediction reported in Ref.~\cite{shen} for protons from 10-20\% central 2.76 TeV \pbpb\ .
} 
\end{figure}

Figure~\ref{alice1} (first row) shows $v_{2i}(p_t)$ data from 2.76 TeV \pbpb\ collisions as reported in Ref.~\cite{alicev2ptb}. This \vtwopt\ vs \pt\ plot format is uninterpretable. A simple transformation to $v_{2i}(p_t) / p_t$ vs $y_{ti}$ (proper transverse rapidity for each hadron species $i$) provides access to a simple description.

Figure~\ref{alice1} (second row) shows transformed ALICE data (points) and a model (dashed) that describes 200 GeV \auau\ minimum-bias (MB) data as reported in Ref.~\cite{quadspec}. Monopole boost $\Delta y_{t0} \approx 0.6$ that defines ratio $p_t' / p_t$ is inferred from zero crossings on $y_{ti}$ for four hadron species.

\section{Deriving quadrupole spectra in the boost frame from $\bf v_2(p_t)$ data in the lab} \label{derive}

\begin{figure}[h]
      \includegraphics[width=1.45in]{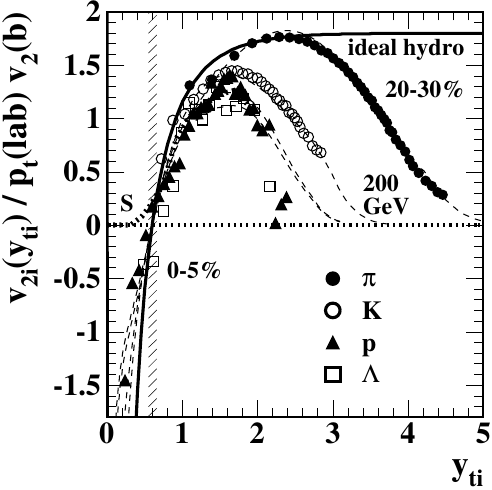}
\includegraphics[width=1.46in]{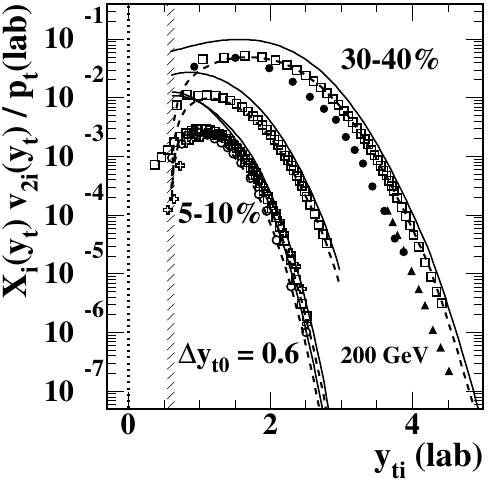}
\put(-128,75) {\bf (a)}
\put(-20,80) {\bf (b)}
    \includegraphics[width=1.45in,height=1.44in]{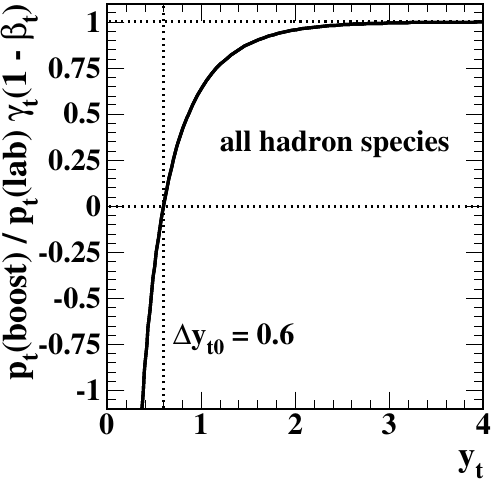}
\includegraphics[width=1.45in]{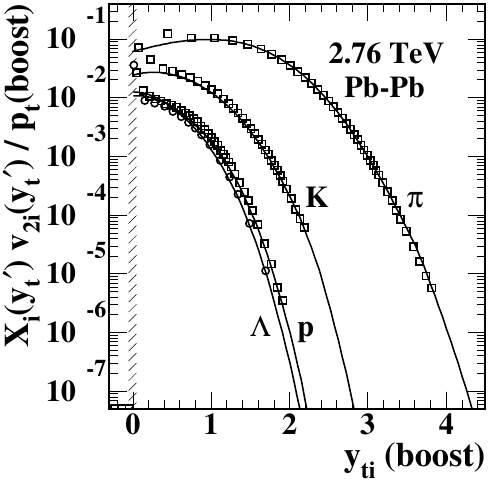}
\put(-128,80) {\bf (c)}
\put(-20,75) {\bf (d)}
\caption{\label{alice8ab}
(a) Data from Fig.~\ref{alice1} (lower) but four species in one plot.
(b) 2.76 TeV data (points) and 200 GeV model (dashed curves) in (a) multiplied by corresponding SP spectra in the form $X_i(y_t)$ for the 30-40\% centrality class of 2.76 TeV \pbpb\ collisions. {The crosses are proton data for 5-10\% central \pbpb.}
The solid dots are 200 GeV pion data from Ref.~\cite{quadspec} rescaled to match \pbpb\ data at lower \yt.
(c) $p_t' / p_t$ for monopole boost $\Delta y_{t0} = 0/6$.
(d) \pbpb\ data and dashed curves from panel (b) divided by ratio $p_t' / p_t$ and shifted left by $\Delta y_{t0} = 0.6$ to the boost frame. The dashed curves transformed from (b) become the solid curves in (d).
}  
\end{figure}

Figure~\ref{alice8ab} shows application of a sequence of transformations as reported in Ref.~\cite{quadspec}. Panel (a) shows data as in Fig.~\ref{alice1} (second row) for four hadron species. Note that proton data (solid triangles) follow the dashed model curves to substantial negative values below a zero crossing on $y_{ti}$ and thus falsify the hydro theory curve S. Panel (b) shows transformation to $X_i(y_{ti}) v_{2i}(y_{ti}) / p_t(\text{lab})$, where $X_i(y_{ti}) = \bar \rho_{0i}(y_{ti}) / \bar \rho_{si}$ (see Sec.~\ref{tcm}). There is one more step to achieve the configuration in Eq.~(\ref{rho2i}).

Figure~\ref{alice8ab} (c) shows ratio $p_t' / p_t$ corresponding to measured value $\Delta y_{t0} = 0.6$. Panel (d) shows panel (b) divided by ratio $p_t' / p_t$ and transformed to the boost frame (i.e.\ shifting data and curves to the left by 0.6) to obtain $X_i(y_{ti}') v_{2i}(y_{ti}') / p_t(\text{boost}) \propto \bar \rho_{2i}(y_{ti}')$ per Eq.~(\ref{rho2i}). The final step is transformation to transverse kinetic energy $m_{ti}' - m_i$ in the boost frame with the proper Jacobian as
\bea
\bar \rho_{2i}(m_{ti}') &\approx&  \left \{ \frac{y_{ti}'}{p_t'(m_{ti}' - m_i)} \right\} \bar \rho_{2i}(y_{ti}') \propto \bar \rho_{0i}(m_{ti}') v_{2i}(m_{ti}') / p_t'.
\eea
The proportionality arises from omitting the constant factors quadrupole boost $\Delta y_{t2}$ and quadrupole slope parameter $T_2$ from Eq.~(\ref{rho2i}).

Assumption of a Boltzmann exponential on \mt\ as a trial function in Eq.~(\ref{combfac}) is validated by data. Evaluation of the integral reveals factor $p_t'$ implicit in quantity $v_2(p_t)$, one reason that the measure is quite misleading. The model assumption is validated via the procedure illustrated in Fig.~\ref{alice8ab} wherein spectrum $\bar \rho_{2i}(y_t)$ emerges as the azimuth average of the integral argument $\bar \rho_{2i}(y_t,\phi_r)$.

\begin{figure}[h]
\begin{center}     \includegraphics[width=3.3in]{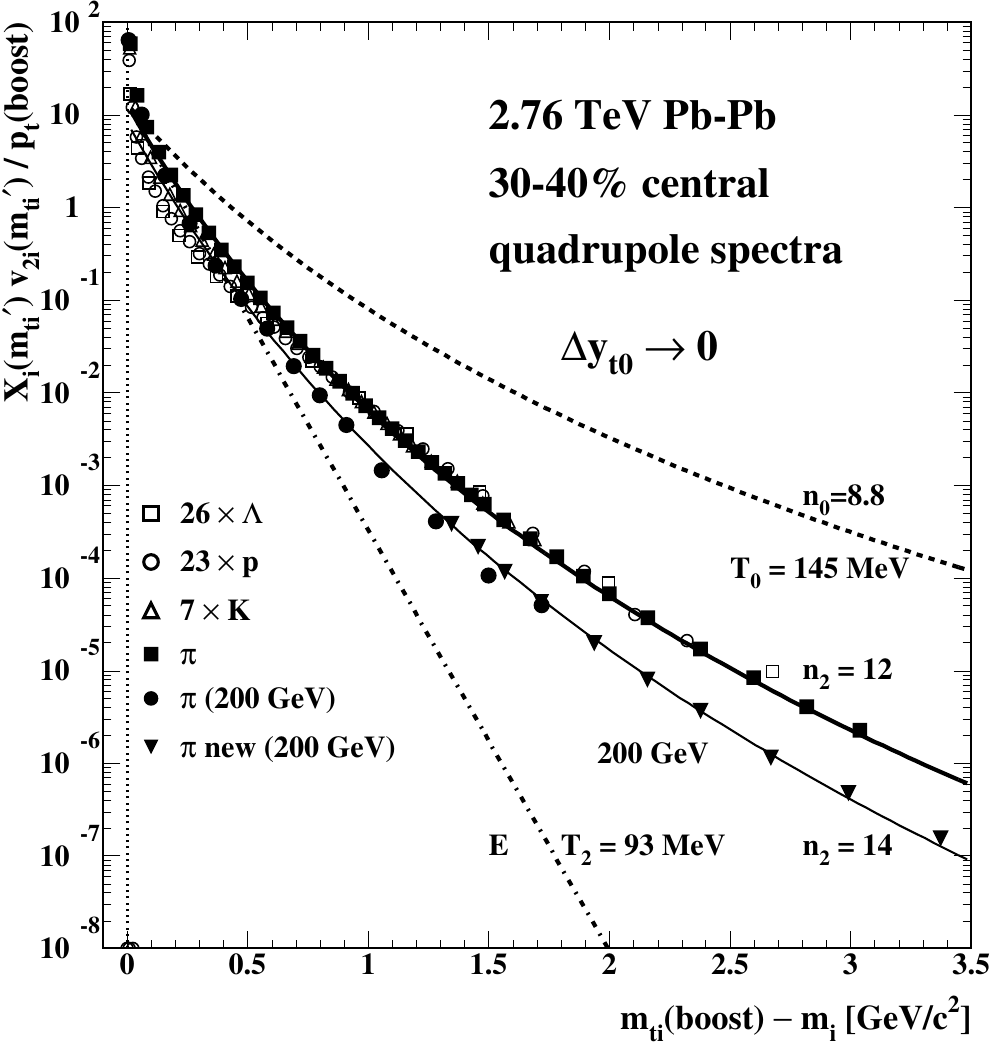}
\put(-80,145) {$\bf \hat S_0(m_t)$}
\put(-99,120) {$\bf \hat S_2(m_t)$}
\end{center}
\caption{\label{alice8c}
Data from Fig.~\ref{alice8ab} (d) transformed  to $m_{ti}' - m_i$ and rescaled (relative to pions) as indicated in the plot. The bold solid curve is model distribution  $\hat S_2(m'_t)$ (Boltzmann exponential) that describes rescaled 2.76 TeV \pbpb\ data. Dash-dotted curve E is the Boltzmann exponential limit for $T_2 = 93$ MeV. The dashed curve is proportional to SP soft component $\hat S_0(m_t)$ for unidentified hadrons from 2.76 TeV \pp\ collisions derived in Ref.~\cite{alicetomspec}. The thin solid curve, solid dots and inverted solid triangles are the 200 GeV pion quadrupole spectrum from Ref.~\cite{quadspec} rescaled to match 2.76 TeV data at low $m_t'$ for shape comparison.
} 
\end{figure}

Figure~\ref{alice8c} shows quadrupole spectra on transverse kinetic energy in the boost frame for four hadron species from 2.76 TeV \pbpb\ collisions as  reported in Ref.~\cite{tomnewquad}. The corresponding quadrupole model ($\hat S_2(m_{ti}')$) is the bold solid curve. Also shown for comparison are quadrupole spectra from 200 GeV \auau\ collisions and corresponding model (thin solid curve). The difference arising with the energy change is entirely due to SP spectrum factors $\bar \rho_{0i}(y_t)$ in Eq.~\ref{rho2i}.

Figure~\ref{alice8c} presents several notable results. (a) Quadrupole spectra for four hadron species (data) coincide within data uncertainties with a simple spectrum model $\hat S_{2}(m_t')$, being a Boltzmann exponential on transverse kinetic energy with a power law tail. The tail should be expected for any real system. (b) Quadrupole spectra are quite distinct from SP spectra, even their soft components $\hat S_{0i}(m_t)$ (e.g.\ dashed curve). (c) Data spectra in this plotting format, based on an {\em extensive} correlation measure (pair numbers), vary on $m_{ti}'$ over seven orders of magnitude but were derived from $v_2(p_t)$ data in Fig.~\ref{alice1} (upper row) that vary over a very limited range. (d) The single bold-solid model curve  $\hat S_{2}(m_t')$ in this figure, back transformed according to the sequence described above, describes $v_{2i}(p_t)$ data for several hadron species over the full \pt\ acceptance of a typical measurement. For instance, see dashed curves in Fig.~\ref{alice1} (lower) and Fig.~\ref{alice8ab}. That comprehensive data description, based on a very simple QCD-based model, begs the question should hydrodynamic theory play any role in a description of the  quadrupole phenomenon?

\section{Viscous hydro theory vs few-gluon interactions}

\begin{figure}[h]
     \includegraphics[width=1.45in]{alicev2-11d}
     \includegraphics[width=1.45in]{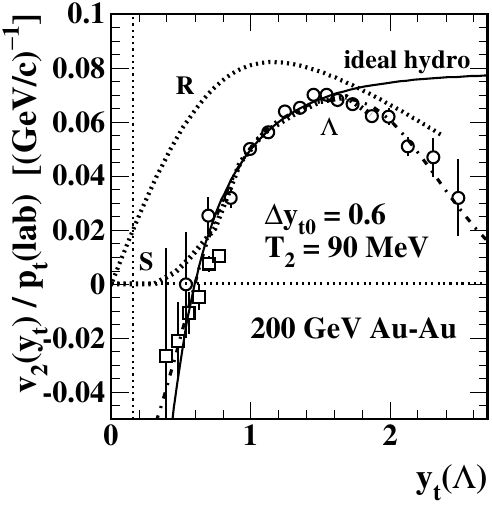}
\put(-130,24) {\bf (a)}
\put(-22,24) {\bf (b)}
     \includegraphics[width=1.45in,height=1.45in]{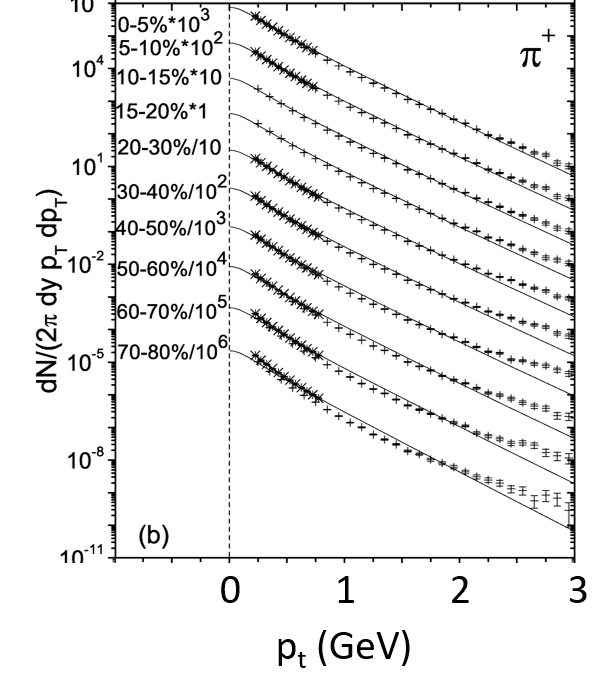}
     \includegraphics[width=1.45in,height=1.5in]{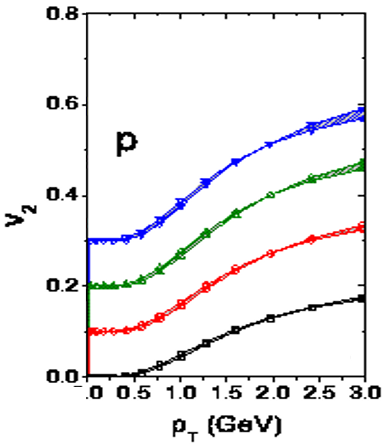}
\put(-139,93) {\bf (c)}
\put(-22,93) {\bf (d)}
\caption{\label{alice1xx}
(a) $v_2(p_t)$ data plotted in a format corresponding to Eq.~(\ref{v2gen}) {as in Fig.~\ref{alice1} (lower). 
The 200 GeV dashed curves in (a) are predictions taken from Ref.~\cite{quadspec}.
Curve S is hydro theory for protons from Ref.~\cite{shen} with upper limit near \yt\ = 1.9.} 
(b) Lambda data from Ref.~\cite{quadspec} plotted on an expanded \yt\ scale compared to older viscous-hydro theory curve R for protons~\cite{rom} {  and more-recent hydro result S (10-20\% central \pbpb)~\cite{shen} also for protons.}
The open boxes are more-recent Lambda data for 0-10\% central \auau\ collisions~\cite{newstarpion}. 
The curve marked ``ideal hydro'' is simply proportional to ratio $p_t'/p_t$ appearing in Fig.~\ref{alice8ab} (c). (c,d) are hydro theory results from Ref.~\cite{shen} including SP \pt\ spectra (c) and \vtwopt\ predictions (d).
} 
\end{figure}

The second and third lines of Eq.~(\ref{combfac}) may be rearranged to obtain
\bea \label{v2gen}
\frac{v_2(p_t,\Delta y_{t0},\Delta y_{t2})}{p_t} &\approx& \frac{p'_t(p_t,\Delta y_{t0})}{p_t} \cdot \frac{ \Delta y_{t2}}{2T_2} \cdot \frac{\bar \rho_2(p_t,\Delta y_{t0})}{\bar \rho_0(p_t)}.
\eea
That relation describes the data in Fig.~\ref{alice1} (lower) and Fig.~\ref{alice1xx} (a,b). If Eq.~(\ref{v2av}) were valid then the ratio of spectrum densities at right in Eq.~(\ref{v2gen}) should cancel and $v_2(p_t)/ p_t \propto p_t' / p_t$ would be true. 

Fig.~\ref{alice1xx} (a) repeats Fig.~\ref{alice8ab} (a) for convenience. The trend $p_t' / p_t$ from Fig.~\ref{alice8ab} (c) marked ``ideal hydro'' has been added for comparison. The data approximately follow that trend at lower \yt\ but then descend rapidly from that trend at higher \yt. The reason is evident given the  known structures of $\bar \rho_2(p_t)$ and $\bar \rho_0(p_t)$ from Refs.~\cite{quadspec,tomnewquad}. At lower \yt\ the cooler exponential in numerator ($T_2 \approx 93$ MeV) compared to denominator ($T_0 \approx 145-200$ MeV) means data rise slower than ratio $p_t' / p_t$. At some point the {\em jet contribution} to $\bar \rho_0(p_t)$ dominates and the ratio falls rapidly away toward zero.

SP spectrum $\bar \rho_{0i}$ with its two distinct components reflecting jet production and participant-nucleon dissociation cannot be  described by a hydro theory that assumes a single dominant {\em monolithic} particle source, as also assumed for Eq.~(\ref{v2av}). That reality is reflected by theory vs data spectrum comparisons in Fig.~\ref{alice1xx} (c) that show large systematic disagreements even over a limited \pt\ interval.

Hydro theory cannot describe \vtwopt\ data except within a narrow \pt\ window ($p_t < 3$ GeV/c) where it is fitted to data. Theory fails to follow data through a zero crossing clearly indicated by data in Fig.~\ref{alice1xx} (a,b) and {\em predicted} by Eq.~(\ref{combfac}). Compare theory trends in panel (d) with data trends in panels (a,b). Theory trends stop at 3 GeV/c and remain {\em at or above zero} down to zero momentum.

Finally, Eq.~(\ref{v2gen}) is able to predict \vtwopt\ data for identified hadrons based on a few fixed parameters as demonstrated above for 2.76 TeV \pbpb\ data. The data description extends over full detector acceptances, not limited to a narrow {\em ad hoc} \pt\ range. The parameter values are determined by SP spectra and jet contributions and quadrupole spectra as in Fig.~\ref{alice8c}. One may then reasonably question {any} relevance for hydro theory regarding high-energy nuclear collisions.

\section{Two-component model (TCM) {\em required} by SP spectrum data} \label{tcm}


As an introduction to the two-component spectrum model a recent analysis of $K^0_\text{S}$ data from Ref.~\cite{alippss} reported in Ref.~\cite{jetspid} is summarized. The result reveals the dominant role of jet production in spectrum evolution. An accurate and {\em interpretable} model for SP \pt\ spectra is essential for several reasons. Within a {\em preferred narrative} applied to heavy ion collisions over the past twenty-five years, spectra are said to ``harden'' with increasing  event multiplicity or centrality, interpreted to arise from radial flow via QGP formation. TCM spectrum analysis conflicts with that conclusion. Accurate spectrum models are required for quadrupole analysis as in Sec.~\ref{derive}. Understanding of spectrum evolution with A-B geometry or event multiplicity is required for proper understanding of spectrum phenomena interpreted to indicate jet modification (quenching) associated  with QGP formation.

Identified hadron (PID) spectra for species $i$ near mid rapidity may be expressed in TCM form as $\bar \rho_{0i}(p_t) = \bar \rho_{si} \hat S_{0i}(p_t) + \bar \rho_{hi}\hat H_{0i}(p_t)$, the sum of soft and hard (jet) components. Functions $\hat X_{0i}(p_t)$ are unit-normal models for soft and hard  shapes derived from spectrum data. Soft-component {\em charge densities} may be expressed as $\bar \rho_{si} = z_{si} \bar \rho_s= \bar \rho_{0i}/(1+\tilde z_i x\nu)$. Rescaled spectra may be defined as $X_i(p_t) = \bar \rho_{0i}(p_t)/\bar \rho_{si} = \hat S_{0i}(p_t) + \tilde z_i x\nu\hat H_{0i}(p_t)$. An expression for  spectrum hard components is $Y_i(p_t) = X_i(p_t) - \hat S_{0i}(p_t) \approx \tilde z_i x\nu\hat H_{0i}(p_t)$. Fractional abundance hard/soft ratios $\tilde z_i \equiv z_{hi} / z_{si}$ are simply proportional to hadron mass~\cite{pidpart1}. The \aa\ hard/soft ratio is $x\nu \equiv \bar \rho_h/\bar \rho_s$ with $\bar \rho_s =(N_{part}/2)\bar \rho_{sNN}$ and $\bar \rho_h = N_{bin }\bar \rho_{hNN}$. The combination $\tilde z_i x\nu$ represents  A-B {\em collision geometry} that should be determined from data, not from a Monte Carlo (e.g.\ Glauber model).

\begin{figure}[h]
	\includegraphics[width=2.95in]{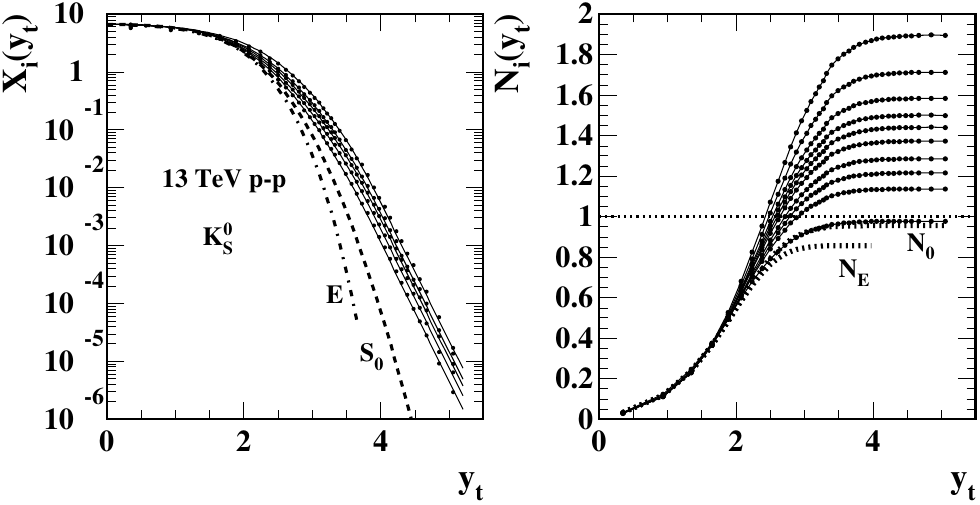}
\put(-125,95) {\bf (a)}
\put(-50,95) {\bf (b)}
	\includegraphics[width=2.95in]{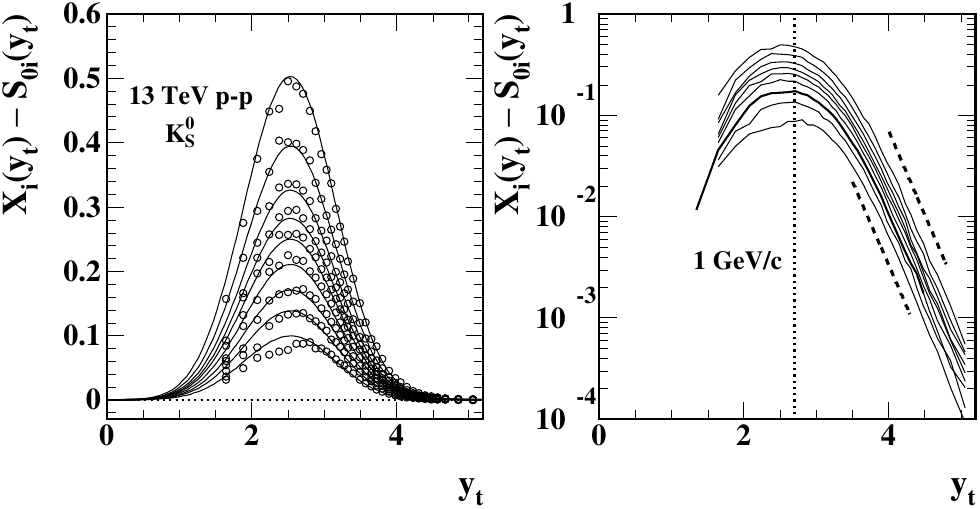}
\put(-125,95) {\bf (c)}
\put(-19,95) {\bf (d)}
	\caption{\label{kosspectra}
(a) Rescaled \k0s\ spectra in the form $X_i(y_t)$ defined above for 13 TeV \pp\ collisions. Solid curves are corresponding TCMs.
(b) Running integrals as defined in Eq.~(\ref{runint}) for data in (a). $N_0(y_t)$ are running integrals of soft components $\hat S_{0i}(y_t)$. 
(c) Spectrum hard components (jet production) in the form $Y_i(y_t)$ defined above.
(d) Data from panel (c) replotted in a semilog format. The dotted line marks $p_t \approx 1$ GeV/c.
}  
\end{figure}

Figure~\ref{kosspectra} (a) shows rescaled spectra $X_i(y_t)$ as above plotted vs transverse rapidity $y_{t} = \ln[(p_t + m_{t})/m]$ (for $\pi$s) that greatly improves visual access to lower-\pt\ structure. Note that $y_t = 1$ is $p_t \approx 0.15$ GeV/c, $y_t = 2$ is $p_t \approx 0.5$ GeV/c, $y_t = 2.7$ is $p_t \approx 1$ GeV/c and $y_t = 4$ is $p_t \approx 4$ GeV/c. Data are shown as solid dots, and solid curves are the TCM from Ref.~\cite{ppsss}. Dashed curves are soft-component models $\hat S_{0i}(y_t)$ from Ref.~\cite{ppsss} -- Boltzmann exponentials on transverse mass $m_{ti}$ (for kaons) with power-law tails. 
Estimation of rescale factors $\bar \rho_{si}$ is quite accurate as indicated by agreement of spectra at low \yt.
Relevance of the TCM details above should be evident below.

Figure~\ref{kosspectra} (b) shows running integrals of the form
\bea \label{runint}
N_{i}(p_t) &\equiv& \int_0^{p_t} dp_t' p_t' X_{i}(p_t')
\\ \nonumber
&\rightarrow& 1 + \tilde z_i x \nu ~~~\text{for $p_t \rightarrow \infty$},
\eea
with $\tilde z_i x\nu =  \bar\rho_{hi}/ \bar \rho_{si}$.
If $X_{i} \rightarrow \hat S_{0i}$ then $N_{i} \rightarrow N_0$ which should  go asymptotically to 1 but falls slightly below due to discrete data integration. The fastest increase of data integrals occurs near \yt\ = 2.7 locating the bulk of jet fragments.
Note that lower-\pt\ parts of running integrals coincide within data uncertainties. Soft components are {\em insensitive} to  the collision space-time volume.

Figure~\ref{kosspectra} (c) shows differences $Y_i(p_t) = X_i(p_t) - \hat S_{0i}(p_t)$ transformed to densities on \yt\ via Jacobian  $p_t m_{ti} / y_{ti}$. Spectrum data are represented by points whereas curves represent the TCM as  in Fig.~\ref{kosspectra} (a). The  amplitudes is depend only on factors $\tilde z_i x \nu$ as defined above. On these linear plots hard components are well approximated by Gaussians on \yt\ suggesting the correct model function {\em as required by data}.
\k0s\ hard components (jets)  persist down to $y_{t\pi} \approx 1.5$ ($p_t \approx 0.3$ GeV/c).

%

Figure~\ref{kosspectra} (d) shows the same results on a semilog plot. 
In this  format exponential (on \yt) tails are clearly evident, as characterized by the dashed lines. An exponential tail on \yt\ is equivalent to a power-law tail on \pt\ that relates to an underlying jet energy spectrum~\cite{fragevo,jetspec2}. 



To summarize, PID spectra for 13 TeV \pp\ collisions are described within data uncertainties {\em over the full detector acceptance} by a few-parameter two-component spectrum model. The TCM is not fitted to individual spectra. Jet fragments are prominent, following expected trends relating to QCD factorization. There is no room for a hydrodynamic description such as the blast-wave model.


\section{A-A geometry inferred from inversion of ensemble-mean \mmpt\ data} \label{geometry}

The spectrum model TCM presented in Sec.~\ref{tcm} requires accurate determination of the A-B collision geometry, taken here to mean not collision centrality but instead the average soft and hard densities per \nn\ collision $\bar \rho_{xNN}$ and the number of nucleon N participant pairs $N_{part}/2$ and the number of \nn\ binary collisions $N_{bin}$. The basis for geometry determination is estimation of hard/soft ratio $x\nu$  from ensemble-mean \mmpt\ data. Ensemble-mean \mmpt\ may be expressed as~\cite{tommpt}
\bea
\bar p_t &=& \frac{\bar P_t}{n_{ch}} = \frac{\bar p_{ts} + x\nu \bar p_{th}}{1 + x\nu},
\eea
where $\bar P_t$ is total \pt\ within $\Delta \eta$ and $\bar p_{tx}$ values may be inferred from TCM descriptions of an assortment of A-B spectra. Thus, given \mmpt\ data ratio $x\nu$ may be inferred by inverting  that relation.

\begin{figure}[h]
	\includegraphics[width=2.95in]{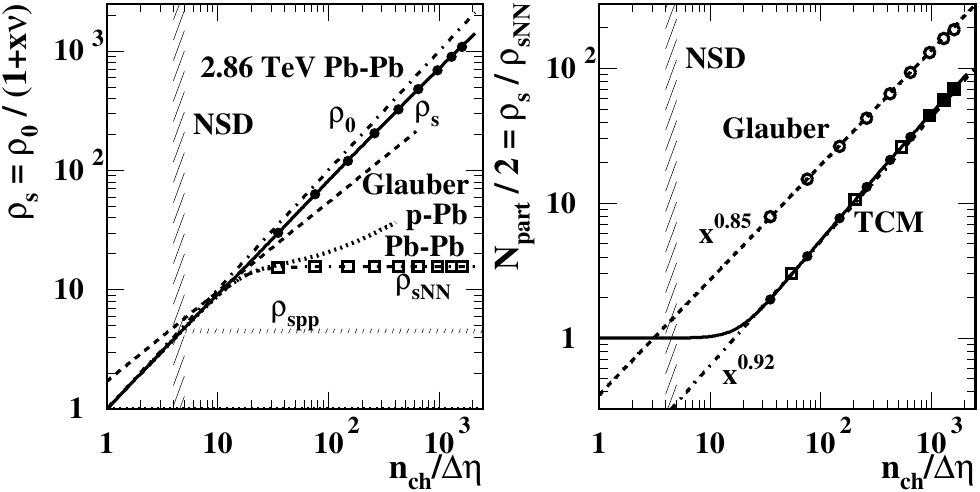}
\put(-132,96) {\bf (a)}
\put(-27,95) {\bf (b)}
	\includegraphics[width=2.95in]{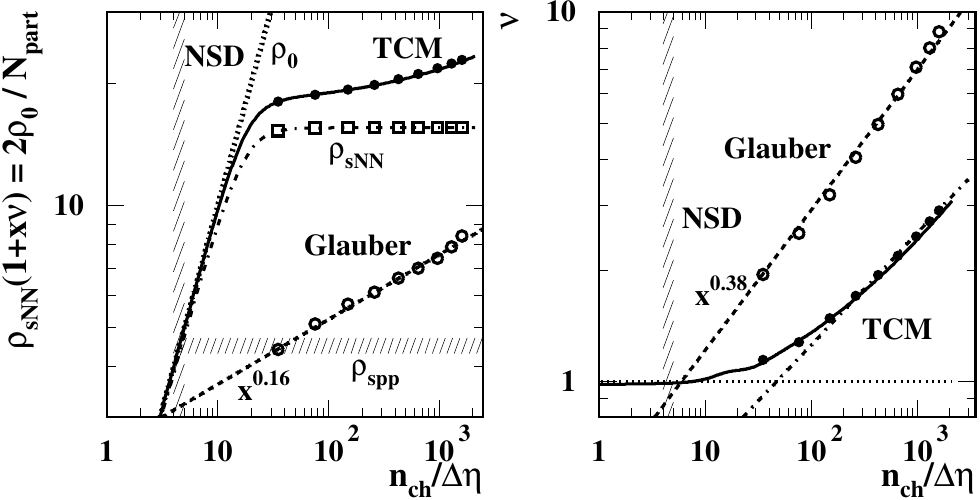}
\put(-145,95) {\bf (c)}
\put(-30,95) {\bf (d)}
	\caption{\label{geom} 		(a) Soft density $\bar \rho_s = \bar \rho_0 / (1+x\nu)$ (solid dots and curve) obtained with TCM $x\nu$ from Ref. ~\cite{tompbpb}. The dash-dotted curve is a $\bar \rho_0$ reference. 
The lower dash-dotted and dotted curves are TCM $\bar \rho_{sNN}$ trends for \pbpb\ and \ppb\ collisions respectively.
		(b) $N_{part}/2 = \bar \rho_s / \bar \rho_{sNN}$ (TCM points and curve) derived from $\bar \rho_s$ (solid) and $x = \alpha  \bar \rho_{sNN}$ conjecture (lower dash-dotted) in (a). Open circles are Glauber values from Ref.~\cite{alicecent}.
		(c) $2 \bar \rho_0 / N_{part} = \bar \rho_{sNN} (1+x\nu)$ for TCM (solid points and curve) and Glauber model (open points and dashed curve).
		(d) Quantity $\nu = 2 N_{bin}/N_{part}$ from the TCM (solid points and curve) and from the Glauber model (open circles). 
For the power laws $x = \bar \rho_0 \equiv n_{ch} / \Delta \eta$.
}  

\end{figure}

Figure~\ref{geom} (a) shows $\bar \rho_s = \bar \rho_0 / (1+x\nu) = \bar \rho_{sNN}(N_{part}/2)$ vs $\bar \rho_0 = n_{ch} / \Delta \eta$ (solid curve) based on an $x\nu$ trend inferred from \pbpb\ \mmpt\ data inversion~\cite{tompbpb}. Based on similar analysis of \ppb\ collisions~\cite{ppbpid,tomglauber} it is likely that for \pbpb\ collisions $\bar \rho_{sNN}$ (open squares) saturates at a constant value above a transition near $\bar \rho_0 = 15$ relating to {\em N-N exclusivity}~\cite{tomexclude}.

Figure~\ref{geom} (b) shows $N_{part} / 2 = \bar \rho_s / \bar \rho_{sNN}$ (solid curve) inferred from trends in panel (a). The trend follows $N_{part} / 2 = 1$ (single peripheral \nn\ interaction) up to a transition point $\bar \rho_0 = 15$ and then increases in participant number as inferred from \mmpt\ data (jet production increase).

Figure~\ref{geom} (c) shows quantity $2\bar \rho_0 / N_{part}$ inferred from panel (b). That result demonstrates the dramatic difference between a {\em published} Glauber Monte Carlo model result (dashed curve and points)~\cite{alicecent} and the trend inferred from physical measurement (solid curve and points)~\cite{tompbpb}. Figure~\ref{geom} (d) shows quantity $\nu \equiv 2N_{part} / N_{bin}$ from which  $N_{bin}$ may be inferred given panel (b).

To summarize, there are major differences between conventional Glauber Monte Carlo estimates and geometry inferred from \mmpt\ data. Dramatic differences arise in part from effects of {\em exclusivity} for \nn\ collisions within A-B collisions~\cite{tomexclude}. The number of participant pairs is typically a factor three lower than the Monte Carlo estimate -- \mmpt\ data are inconsistent with a high-density system.

\section{Extended PID spectrum TCM applied to 2.76 TeV Pb-Pb collisions}

Conventional spectrum analysis relating to claims of ``jet quenching'' as a signal of QGP formation focuses on spectrum ratio $R_\text{AA}$ defined as 
\bea
R_\text{AA}(p_t) &\equiv& \frac{1}{N_{bin}}\frac{\bar \rho_\text{0AA}(p_t)}{\bar \rho_{0pp}(p_t)}.
\eea
The measure so defined is expected to apply above a \pt\ value seen as a {\em lower bound for jet fragments}, say 4 GeV/c. Within that interval deviations below unity are interpreted to indicate jet quenching.

There are several severe problems with that measure definition and its interpretation. (a) The two spectra in the ratio are demonstrably composite, as described by the TCM and demonstrated in detail in Sec.~\ref{tcm}. Within the ratio the hard component prevails at higher \pt\ while the soft component prevails at lower \pt. But the great majority of jet fragments appears within a peak with mode near 1 GeV/c ($y_t \approx 2.7$) as in Fig.~\ref{kosspectra} (c). Ratio $R_\text{AA}$ then {\em hides more than 90\%} of the produced jet fragments. (b) $N_{bin}$ is conventionally estimated with a Glauber Monte Carlo. The negative consequences may be viewed in Sec.~\ref{geom}. (c) The assumption that jet production should scale with $N_{bin}$, referring to nucleon \nn\ collisions, is unwarranted and has been falsified in Rev.~\cite{tompbpb}.

An alternative approach involves intact spectrum hard components $Y_i(p_t) = X_i(p_t) - \hat S_{0i}(p_t)$ without any rescaling. A replacement for $R_\text{AA}$ is then defined for \pbpb\ collisions as
\bea \label{ryy}
r_Y(y_t) &=&  \frac{Y_{PbPb}(y_t)}{Y_{pp}(y_t)} \approx \frac{\tilde z_i x \nu_{PbPb} \hat H_{0PbPb}(y_t)}{\tilde z_i x_{NN} \hat H_{0NN}(y_t)}.
\eea
The numerator and denominator include the complete hard components (all jet fragments) for the two collision systems. The denominator is a TCM NN model to compare with \pp\ data. 

\begin{figure}[h]
	\includegraphics[width=1.45in]{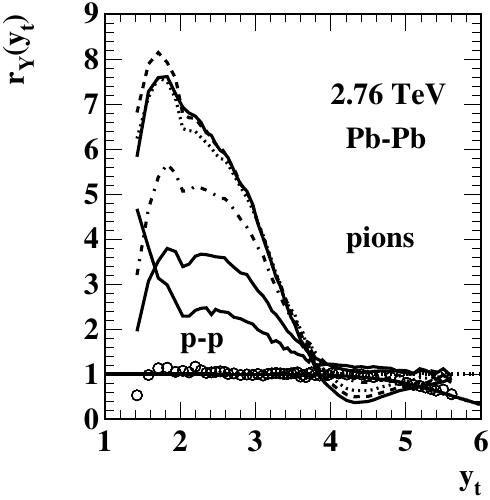}
	\includegraphics[width=1.45in]{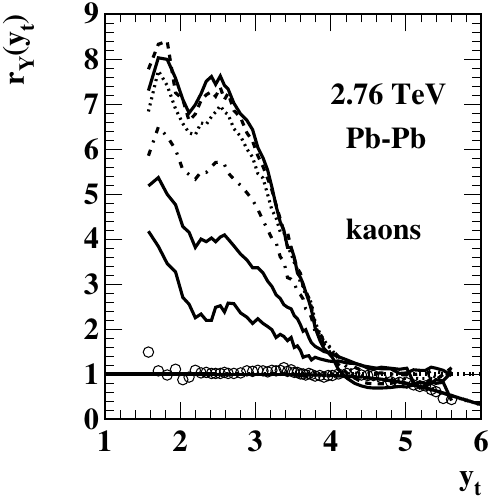}
\put(-130,92) {\bf (a)}
\put(-22,92) {\bf (b)}
	\includegraphics[width=2.95in]{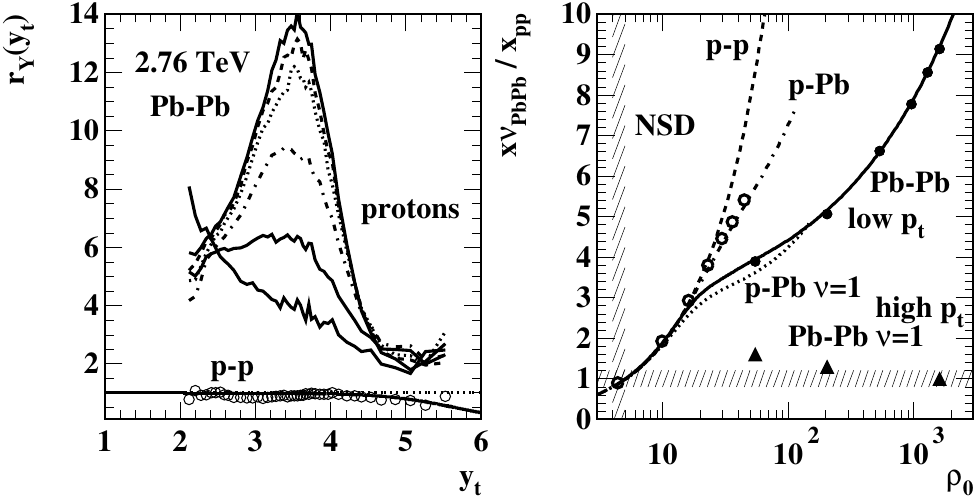}
\put(-128,92) {\bf (c)}
\put(-27,92) {\bf (d)}
	\caption{\label{ry}
Spectrum ratios $r_{Y}(y_t)$ for three hadron species from 2.76 TeV \pbpb\ collisions (a,b,c) showing reduction to {\em unity} above \yt\ = 4 ($p_t \approx 4$ GeV/c). (d) ratio values (solid curve) that appear in (a,b,c) at {\em lower} \yt\ assuming there is no jet modification and hard components are factorizable. Note that \pp\ spectra from Ref.~\cite{alicepbpbpidspec} (open circles) coincide with the \nn\ model up to a point and then descend (lowest solid curves). The \nn\ model is based on several sources of \pp\ spectrum data. Such a deviation is thust unexpected.
}  
\end{figure}

Figure~\ref{ry} (a,b,c) shows ratios $r_Y(y_t)$ for three hadron species from 2.76 TeV \pbpb\ collisions~\cite{alicepbpbpidspec}. In the ratio of Eq.~(\ref{ryy}) factors $\tilde z_i$ cancel for a given hadron species. Assuming the hard-component model functions $\hat H_{0xx}(y_t)$ are equivalent the ratio value should vary with collision multiplicity as Fig.~\ref{ry} (d) (solid curve). The data trends at {\em lower} \yt\ approximately follow that trend.

As reported in Ref.~\cite{tompbpb} hard components $x \nu_{AB} \hat H_{0AB}(y_t)$ cannot be factorized. What matters is not \nn\ interactions but {\em partonic} interactions within \nn\ collisions. Experimentally, {\em exclusivity and time dilation} apply to partons rather than nucleons. Exclusivity says that a nucleon may interact (via partons) with only one nucleon {\em at a time}. For low-$x$ (slow) gluons clocks run fast and a nucleon may interact again quickly. For high-$x$ gluons clocks run slow (time dilation) and a nucleon may be frozen for {\em more than a nucleus diameter}.  Gluon $x$ translates to fragment hadron \pt, and the fragment yield is thereby modulated accordingly as in Fig.~\ref{ry}. Higher-\yt\ hadrons effectively emerge from single \nn\ ($\approx$ \pp) collisions within \aa\ collisions. There is no attenuation by a medium.

\section{Model fits to 2D angular autocorrelations}

2D angular {\em autocorrelation} provide the proper basis for accurate quantitative measurement of two-particle angular correlations. ``Autocorrelation'' is the correct term for this centrally important statistical measure, not a neologism assigned to statistical bias arising from {\em not} using an autocorrelation approach. Angular autocorrelations may be displayed as histograms of {\em pair densities} on {\em difference variables} $(\eta_\Delta,\phi_\Delta)$, where $x_{\Delta} \equiv x_i - x_j$ for particles $i$ and $j$. The primary {\em extensive measure} of correlation amplitudes is then numbers of correlated pairs.

\begin{figure}[h]
	\includegraphics[width=1.45in]{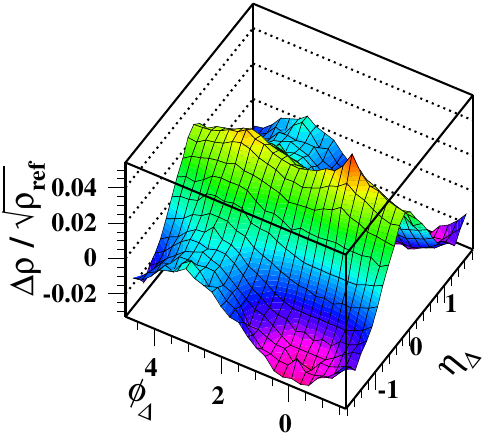}
\put(-20,87) {\bf (a)}
	\includegraphics[width=1.45in]{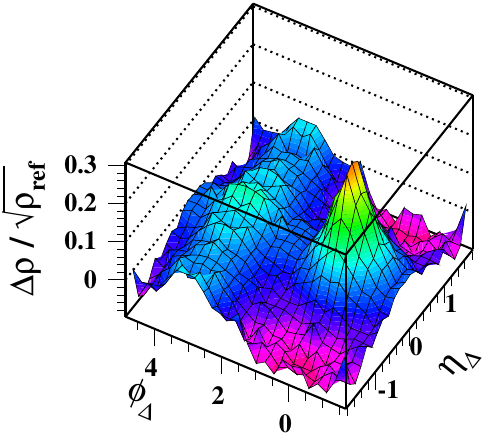}
\put(-20,87) {\bf (b)}
	\includegraphics[width=1.45in]{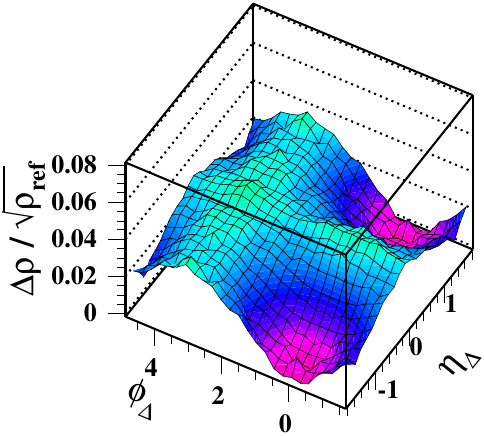}
\put(-20,87) {\bf (c)}
	\includegraphics[width=1.45in]{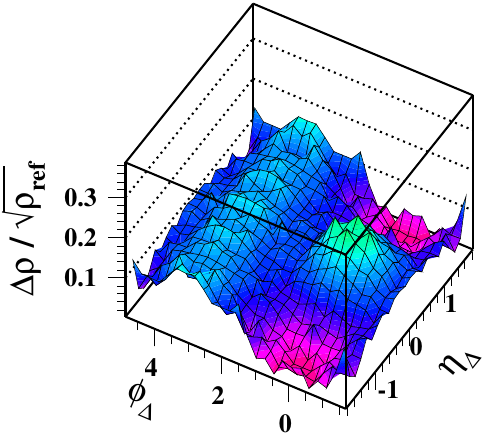}
\put(-20,87) {\bf (d)}
	\caption{\label{modelfits}
Left: 2D angular autocorrelations from 200 GeV \pp\ collisions for low (a) and high (b) multiplicity. Right: Jet and nonjet-quadrupole correlations after subtracting certain fit model elements for low (c) and high (d) event multiplicity.
}  
\end{figure}

Figure~\ref{modelfits} (a,b) shows 2D angular autocorrelations from 200 GeV \pp\ collisions~\cite{ppquad} as pair densities on difference variables for \pp\ event classes 1 (a) and 6 (b) (of 7). The measured quantity is denoted by $\Delta \rho/\sqrt{\rho_{ref}}$. 
A six-element fit model applied to the measured histograms achieves data descriptions within point-point data uncertainties~\cite{ppquad}. 2D  fit residuals are consistent with statistical errors. Model amplitudes are denoted by $A_\text X$ as in $A_\text{2D}$ for  same-side 2D jet peak, $A_\text D$ for away-side jet dipole and $A_\text{Q}$ for nonjet quadrupole. Conversion to pair numbers requires added factor $\bar \rho_0$.

Figure~\ref{modelfits} (c,d) shows data from (a,b) with model elements describing projectile-nucleon dissociation (1D Gaussian on $\eta_\Delta$), Bose-Einstein correlations (narrow 2D exponential peak at the origin) and constant offset subtracted. Remaining structure represents minimum-bias jet production (same-side 2D peak, away-side dipole $\cos(\phi_{\Delta}-\pi)$) and a nonjet quadrupole element represented by $\cos(2\phi_\Delta)$. In reference to Sec.~\ref{poskvol} no reaction plane is necessary, even if one actually existed.

The structures in Fig.~\ref{modelfits} (c,d) may be interpreted as follows. In panel (c) (lowest multiplicity) the jet contribution dominates. The same-side peak (intra-jet correlations) is broad an azimuth $\phi_\Delta$ and the away-side dipole (back-to-back jets) dominates. In panel (d) the quadrupole $\cos(2\phi_\Delta)$ contribution becomes evident by superposition of its two lobes on azimuth. The away-side lobe now reduces the radius of curvature of the away-side dipole peak dramatically. The same-side 2D jet peak appears narrower on azimuth, but that is a result of the same-side lobe of the quadrupole structure superimposed. The curvatures at larger $\eta_\Delta$ near $\phi_\Delta = 0$ respond consistently. At low multiplicity the structure there has a strong minimum (jet dipole). At higher multiplicity the curvature is approximately zero -- filled in by the quadrupole same-side lobe contribution.

\section{Predicting color multipole amplitude trends for A-B collision systems}

Reference~\cite{jetspid} presents a unified description of jet production and quadrupole production within A-B collisions based on QCD color multipole radiation of hadrons from few-gluon interactions, with dipole radiation (dijets) arising from two-gluon interactions and quadrupole radiation arising from three-gluon interactions. This section describes {\em predictions} for amplitude (number of correlated pairs) variation of color multipoles based on A-B collision geometry.

\subsection{Color quadrupole $V_2^2$ trend} \label{quadrupole}

Reference~\cite{ppquad} reports quadrupole amplitudes (pair numbers) for 200 GeV \pp\ collisions in the form $V_2^2\{\text{2D}\} \equiv \bar \rho_0 A_\text{Q}\{2\text D\}$ (solid triangles) vs $\bar \rho_s$, where \{2D\} denotes data inferred from 2D model fits vs other $v_2$ ``methods.'' $\bar \rho_s$ is the soft-component density in $\bar \rho_0 = \bar \rho_s + \bar \rho_h$ which may be interpreted as proportional to the density of low-$x$ gluons.  \pp\ data reveal that the quadrupole amplitude increases as  $V_2^2 \propto \bar \rho_s^3$, suggesting that the quadrupole arises from {\em three-gluon interactions.}

\begin{figure}[h]

\includegraphics[width=1.45in]{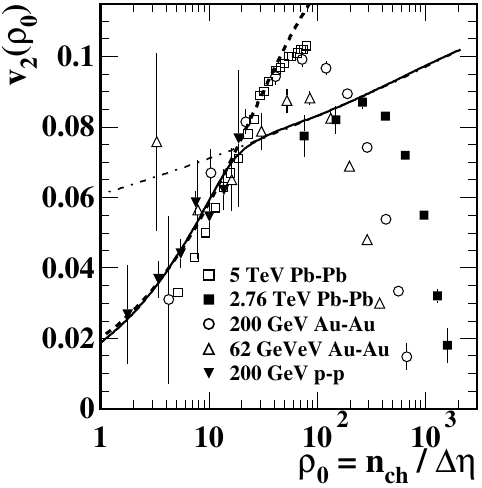}
\includegraphics[width=1.45in]{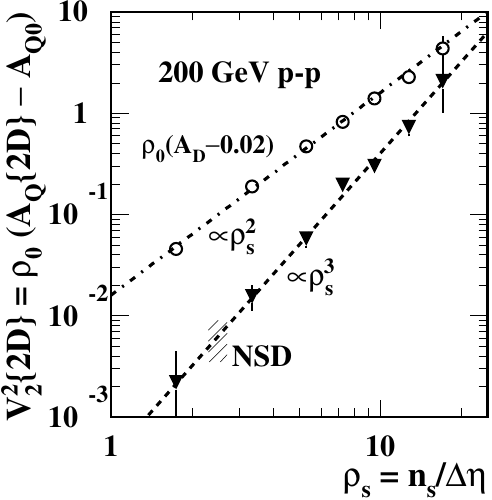}
\put(-124,79) {\bf (a)}
\put(-18,65) {\bf (b)}
	\includegraphics[width=1.45in]{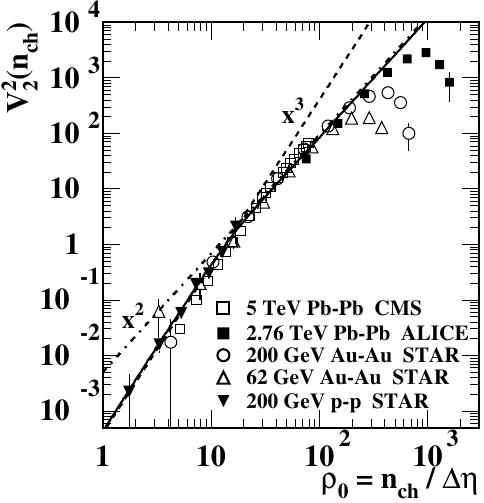}
\includegraphics[width=1.45in]{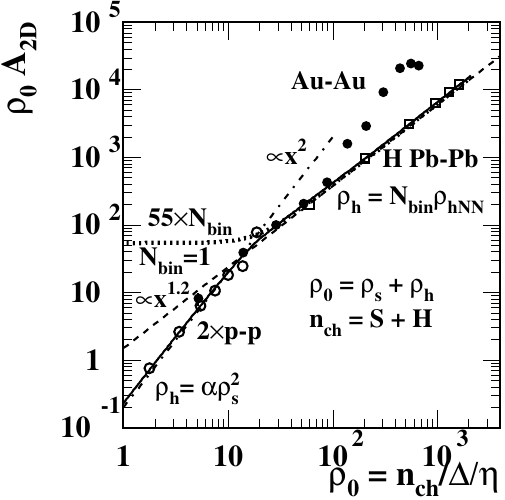}
\put(-126,65) {\bf (c)}
\put(-20,24) {\bf (d)}
	\caption{\label{powerlaw}
	(a) The contents of  (c) transformed to conventional measure $v_2(\bar \rho_0) = \sqrt{V_2^2(\bar \rho_0)} / \bar \rho_0$.
	(b) Total correlated pairs for 200 GeV \pp\ quadrupole trend $V_2^2\{\text{2D}\}$ (solid triangles) and away-side dipole 1D jet peak $\bar \rho_0 A_D$ (open circles). 
	(c)  Quadrupole correlations from 62 and 200 GeV \auau\ collisions (open circles and triangles) from Ref.~\cite{anomalous} and 2.76 TeV \pbpb\ collisions (solid squares) from Ref.~\cite{alicev2b} as  $V^2_2(\bar \rho_0)$ vs charge density $\bar \rho_0$. {Also included are 5 TeV \pbpb\  data (open squares) from Ref.~\cite{cmsabv2}.}  200 GeV \pp\ data from (b) (solid triangles) are superimposed with their cubic trend (dashed). The solid curve is Eq.~(\ref{rhosrhoh})  rescaled to best match \pp\ and \auau\ data at lower \nch. The dash-dotted line $\propto \bar \rho_0^2$ is explained in the text.	
	(d)  Dijet amplitude $\bar \rho_0 A_{2D}$ (2D peak volume) from 200 GeV \auau\ collisions (solid dots) and 2.76 TeV \pbpb\ collisions (open squares)  vs charge density $\bar \rho_0$. \pp\ $\bar \rho_0 A_\text D$ data (open circles) and trend (dash-dotted line) are derived from panel (b).
The solid curve is Eq.~(\ref{v22bxx}) rescaled to best match \pp\ and \auau\ data at lower \nch. 
}  
\end{figure}

Figure~\ref{powerlaw} (a) shows quadrupole data for several A-B systems plotted in the conventional format $v_2(\bar \rho_0)$ vs $\bar \rho_0$ with $\bar \rho_0$ and $v_2$ as published.  ALICE 2.76 TeV \pbpb\ $v_2(n_{ch})$ data are from Ref.~\cite{alicev2b}, ALICE $\bar \rho_0(n_{ch})$ data are from Ref.~\cite{alicerho0}, STAR \auau\  $\bar \rho_0(n_{ch})$ data are from Tables~III and IV of Ref.~\cite{anomalous} and STAR \auau\ $v_2(n_{ch})$ and $A_\text{Q}(n_{ch})$ data are from Ref.~\cite{v2ptb}. 
CMS 5 TeV \pbpb\ $v_2(n_{ch})$ data are from Ref.~\cite{cmsabv2}.
200 GeV \pp\ data and cubic trend (dashed line) are derived from panel (a) via $v_2(\bar \rho_0) = \sqrt{V_2^2(\bar \rho_0)} / \bar \rho_0$. It is reasonable to assert that this presentation of \vtwopt\ data is uninterpretable, although it may serve as a projection screen relating to confirmation bias.

Figure~\ref{powerlaw} (b) shows the quadrupole amplitude (solid triangles, dashed line) for 200 GeV \pp\ collisions in the form $V_2^2\{\text{2D}\} = \bar \rho_0 A_Q \propto \bar \rho_s^3$ (number of pairs) derived via model fits to 2D angular autocorrelations as in Fig.~\ref{modelfits} (c,d) vs soft-component charge density $\bar \rho_s$.
Since for \pp\ collisions $\bar \rho_h \propto \bar \rho_s^2$, the \pp\ quadrupole trend might also be expressed as $V_2^2(n_s) \propto \bar \rho_s \bar \rho_h$. Is that expression universal? For general A-B collisions the product $\bar \rho_s \bar \rho_h$ may be expressed as
\bea \label{rhosrhoh}
\bar \rho_s \bar \rho_h \rightarrow \bar \rho_{sNN} (N_{part}/2) \times \bar \rho_{hNN} N_{bin},
\eea 
which requires accurate determination of A-B collision geometry. From figures in Sec.~\ref{geom} one observes a consistent transition near $\bar \rho_0 = 15$ in A-A collisions. Single peripheral \nn\ collisions below that point transition to certain power-law trends above that point. Below the transition $N_{part}/2 = N_{bin} = 1$. Above the transition $N_{part}/2 \propto \bar \rho_0^{0.92}$, $N_{bin} \approx (N_{part}/2)^{4/3} \propto \bar \rho_0^{1.2}$ and $\bar \rho_{hNN} \propto \bar \rho_{sNN}^2$ in all cases. Those trends may be combined to a single function per Eq.~(\ref{rhosrhoh}) applied to $V_2^2$ data from a range of collision systems from 200 GeV \pp\ to 5 TeV \pbpb.

Figure~\ref{powerlaw} (c) shows quadrupole data for A-B systems from panel (a) plotted as $V_2^2 \equiv \bar \rho_0^2 v_2^2$ with $\bar \rho_0$ and $v_2$ as published, {\em with no rescaling}.  
200 GeV \pp\ data and cubic trend (solid triangles) are copied from panel (b).
Quadrupole  amplitude $V_2^2(\bar \rho_0)$ (correlated pairs) increases by {\em six orders of magnitude} consistently across five collision systems and is described accurately by Eq.~(\ref{rhosrhoh}) (solid curve). Data falloff at high $\bar \rho_0$ may be due to eventwise {\em multiple} three-gluon interactions randomly oriented on azimuth  thus partially canceling. The dash-dotted line is Eq.~(\ref{rhosrhoh})  above $\bar \rho_0 \approx 15$.

\subsection{Color dipole $\bar \rho_0 A_D$ or $\bar \rho_0 A_{2D}$ (dijet) trend} \label{dipole}


Figure~\ref{powerlaw} (b) also shows the amplitude of the away-side dipole 1D jet peak (pair numbers) $\bar \rho_0 A_\text D$ (open circles, dash-dotted line, analogous to $V_2^2$) for 200 GeV \pp\ collisions from Ref.~\cite{ppquad}. 
The \pp\ dipole amplitude increases $\propto \bar \rho_s^2$ as expected from SP spectrum analysis with $\bar \rho_h \propto \bar \rho_s^2$~\cite{ppprd}, consistent with dijets as color dipoles resulting from two-gluon interactions. 

By analogy with Sec.~\ref{quadrupole} one may conjecture that dijet trends for \aa\ collisions should then vary as $\bar \rho_h \rightarrow N_{bin} \bar \rho_{hNN}$ with $\bar \rho_{hNN} \approx \alpha \bar \rho_{sNN}^2$. The corresponding equivalent to the expression for $V_2^2(n_s)$ Eq.~(\ref{rhosrhoh}) would then be
\bea \label{v22bxx}
\bar \rho_0 A_\text{D}(n_{s}) &\propto&\bar \rho_0 A_\text{2D}(n_{s})  \propto \bar \rho_h \propto  N_{bin} \ \bar \rho_{sNN}^2,
\eea
where $\bar \rho_0 A_\text{2D}$ is the same-side 2D jet peak amplitude.

Figure~\ref{powerlaw} (d) shows quantity $\bar \rho_0 A_\text{2D}$ reflecting {total number of correlated pairs} as an {extensive} correlation measure (points). Solid dots are 200 GeV \auau\ data from Ref.~\cite{anomalous}.  Open squares are integrated hard-component yields $H$ obtained from 2.76 TeV \pbpb\ $\pi$ spectra reported in Ref.~\cite{alicepbpb} rescaled by a single factor to match the solid curve. 
 Those \aa\ data correspond to $\bar \rho_0 A_\text D$ data for 200 GeV \pp\ collisions, both representing dijet production. The \pp\ dash-dotted line and open circles are from panel (b).  The solid curve represents density $\bar \rho_h$ defined by Eq.~(\ref{v22bxx}) with parameters  from Sec.~\ref{geom}. Jet amplitudes increase more  than four orders of magnitude across A-B systems.

\section{Summary}

Several points emerge from analyses in Refs.~\cite{tomnewquad,tompbpb,jetspid}:
(i) The assumption of a lone particle source (i.e.\ QGP) for ratio \vtwopt\ motivated by extrapolation from Bevalac flow results is wrong.
(ii) Single-particle (SP) spectra are dominated at lower \pt\ by jets, not flows.
(iii) There is evidence against a dense QCD medium or ``jet quenching.''
(iv) There is evidence against  relevance for hydro theory models.
(v) Few-gluon interactions  are fundamental within A-B collisions.
\pp\ 2D angular correlations plus accurate {\em jet-based} A-B geometry {\em predict}  color quadrupole radiation trends over six orders of magnitude and color dipole radiation (jet) trends over four orders of magnitude.

Those findings resolve several chronic issues after twenty-five years of heavy ion research at the RHIC and LHC. (a) Claims of QGP formation as a ``perfect liquid'' were in conflict with observation of copious low-energy jets and low-\pt\ jet fragments already in the first years of RHIC operation. (b) Systematic trends for \vtwopt\ amplitude variation did not correspond to an observed ``sharp transition'' of jet characteristics for \auau\ collisions. Claims of radial flow as the source of ``spectrum hardening'' conflicted with observations of a dominant jet contribution to spectra. And observation of phenomena in small collision systems at the LHC that had been previously assigned as unique ``signals'' of QGP formation conflicted with the motivating argument for a heavy ion collider, to generate large energy and matter densities for QGP formation. As related in this presentation, all A-B collision systems are simply described based on elementary QCD phenomena.


\begin{thebibliography}{99}

\bibitem{tomnewquad} T.~A.~Trainor,
``Quadrupole spectra derived from 2.76-TeV Pb-Pb identified-hadron v2(pt) data,''
Phys. Rev. C \textbf{114}, no.3, 034911 (2026),
arXiv:2512.08003 [hep-ph].

\bibitem{tompbpb} T.~A.~Trainor,
``Space-time geometry of small and large collision systems at LHC energies,''
arXiv:2510.05314 [hep-ph].

\bibitem{jetspid} T.~A.~Trainor,
``Few-gluon interactions and multipole radiation in high energy nuclear collisions,''
arXiv:2607.25017 [hep-ph].

\bibitem{sphere}  P.~Danielewicz and M.~Gyulassy,
``Jacobian free global event analysis,''
Phys. Lett. B \textbf{129}, 283-288 (1983).

\bibitem{olli} J.~Y.~Ollitrault,
``Anisotropy as a signature of transverse collective flow,''
Phys. Rev. D \textbf{46}, 229-245 (1992).

\bibitem{poskvol} A.~M.~Poskanzer and S.~A.~Voloshin,
``Methods for analyzing anisotropic flow in relativistic nuclear collisions,''
Phys. Rev. C \textbf{58}, 1671-1678 (1998),
arXiv:nucl-ex/9805001 [nucl-ex].

\bibitem{quadspec}   T.~A.~Trainor,
  ``Azimuth quadrupole component spectra on transverse rapidity ${\bf y_t}$ for identified hadrons from Au-Au collisions at $\sqrt{s_{NN}} =$ 200 GeV,''
  Phys.\ Rev.\  C {\bf 78}, 064908 (2008),
arXiv:0803.4002 [hep-ph].

\bibitem{alicev2ptb} B.~B.~Abelev \textit{et al.} [ALICE],
``Elliptic flow of identified hadrons in Pb-Pb collisions at $ \sqrt{s_{\mathrm{NN}}}=2.76 $ TeV,''
JHEP \textbf{06}, 190 (2015),
arXiv:1405.4632 [nucl-ex].

\bibitem{shen} C.~Shen, U.~Heinz, P.~Huovinen and H.~Song,
``Radial and elliptic flow in Pb+Pb collisions at the Large Hadron Collider from viscous hydrodynamic,''
Phys. Rev. C \textbf{84}, 044903 (2011),
arXiv:1105.3226 [nucl-th].

\bibitem{alicetomspec}T.~A.~Trainor,
``Charge-multiplicity and collision-energy dependence of $p_t$ spectra from $p$-$p$ collisions at the relativistic heavy-ion collider and large hadron collider,''
J.\ Phys.\ G \textbf{44}, no.7, 075008 (2017),
arXiv:1603.01337 [hep-ph].

\bibitem{rom} P.~Romatschke and U.~Romatschke,
  ``Viscosity Information from Relativistic Nuclear Collisions: How Perfect is
  the Fluid Observed at RHIC?,''
  Phys.\ Rev.\ Lett.\  {\bf 99}, 172301 (2007),
arXiv:0706.1522 [nucl-th].

\bibitem{newstarpion} B.~I.~Abelev {\it et al.} (STAR Collaboration),
``Centrality dependence of charged hadron and strange hadron elliptic flow from s(NN)**(1/2) = 200-GeV Au + Au collisions,''
Phys.\ Rev.\ C {\bf 77}, 054901 (2008),
arXiv:0801.3466 [nucl-ex].

\bibitem{alippss} S.~Acharya \textit{et al.} (ALICE),
``Multiplicity dependence of (multi-)strange hadron production in proton-proton collisions at $\sqrt{s}$ = 13 TeV,''
Eur.\ Phys.\ J. C \textbf{80}, no.2, 167 (2020),
arXiv:1908.01861 [nucl-ex].

\bibitem{pidpart1} T.~A.~Trainor,
``Precision identified-hadron spectrum analysis for 5 TeV $\bf p$-$\bf Pb$ collisions -- Part I,''
arXiv:2112.09790 [hep-ph].

\bibitem{ppsss}
T.~A.~Trainor,
``Systematic analysis of (multi)strange hadron $\bf p_t$ spectra from small collision systems at the large hadron collider,''
arXiv:2303.14299 [hep-ph].
	
\bibitem{fragevo}    T.~A.~Trainor,
``Evolution of minimum-bias parton fragmentation in nuclear collisions,''
Phys.\ Rev.\  C {\bf 80}, 044901 (2009),
arXiv:0901.3387 [hep-ph].

\bibitem{jetspec2}  T.~A.~Trainor,
``Universal parametrization of jet production based on parton and fragment rapidities,''
Phys.\ Rev.\ D  {\bf 89}, 094011 (2014),
  arXiv:1403.3685 [hep-ph].

\bibitem{tommpt} T.~A.~Trainor,
``Ensemble-mean $\bf p_t$ and hadron production in high-energy nuclear collisions,''
arXiv:1708.09412 [hep-ph].



\bibitem{tomglauber}  T.~A.~Trainor,
''Glauber-model analysis of 5 TeV $\bf p$-Pb centrality compared to a two-component (soft + hard) model of hadron production in high-energy nuclear collisions''
arXiv:1801.05862 [hep-ph].

\bibitem{ppbpid}  T.~A.~Trainor,
``A two-component model for identified-hadron $\bf p_t$ spectra from 5 TeV p-Pb collisions,''
J.\ Phys.\ G \textbf{47}, no.4, 045104 (2020),
arXiv:1812.01151 [hep-ph].

\bibitem{tomexclude}  T.~A.~Trainor,
``Exclusivity of p-N interactions within p-A collisions,''
arXiv:1801.06579 [hep-ph].

\bibitem{alicecent}  K.~Aamodt {\it et al.} (ALICE Collaboration),
``Centrality dependence of the charged-particle multiplicity density at mid-rapidity in Pb-Pb collisions at $\sqrt{s_{NN}}=2.76$ TeV,''
Phys.\ Rev.\ Lett.\  {\bf 106}, 032301 (2011),
arXiv:1012.1657 [nucl-ex].

\bibitem{alicepbpbpidspec}  J.~Adam {\it et al.} (ALICE Collaboration),
``Centrality dependence of the nuclear modification factor of charged pions, kaons, and protons in Pb-Pb collisions at $\sqrt{s_{\rm NN}}=2.76$ TeV,''
Phys.\ Rev.\ C {\bf 93}, no. 3, 034913 (2016),
arXiv:1506.07287 [nucl-ex].

\bibitem{ppquad}  T.~A.~Trainor and D.~J.~Prindle,
 ``Charge-multiplicity dependence of single-particle transverse-rapidity $y_t$ and pseudorapidity $\eta$ densities and 2D angular correlations from 200 GeV p-p collisions,''
  Phys.\ Rev.\ D {\bf 93}, 014031 (2016),
arXiv:1512.01599 [hep-ph].

\bibitem{anomalous}  G.\ Agakishiev, {\it et al.} (STAR Collaboration),
``Anomalous centrality evolution of two-particle angular correlations from Au-Au collisions at $\sqrt{s_{\rm NN}}$ = 62 and 200 GeV,''
  Phys.\ Rev.\ C {\bf 86}, 064902 (2012),
arXiv:1109.4380 [nucl-ex].

\bibitem{alicev2b}  K.~Aamodt {\it et al.} (ALICE Collaboration),
``Elliptic flow of charged particles in Pb-Pb collisions at 2.76 TeV,''
Phys.\ Rev.\ Lett.\  {\bf 105}, 252302 (2010),
arXiv:1011.3914 [nucl-ex].

\bibitem{cmsabv2} A.~M.~Sirunyan \textit{et al.} (CMS),
``Observation of Correlated Azimuthal Anisotropy Fourier Harmonics in $pp$ and $p+Pb$ Collisions at the LHC,''
Phys.\ Rev.\ Lett.\ \textbf{120}, no.9, 092301 (2018),
arXiv:1709.09189 [nucl-ex].

\bibitem{alicerho0} K.~Aamodt \textit{et al.} (ALICE),
``Centrality dependence of the charged-particle multiplicity density at mid-rapidity in Pb-Pb collisions at $\sqrt{s_{NN}}=2.76$ TeV,''
Phys.\ Rev.\ Lett.\ \textbf{106}, 032301 (2011),
arXiv:1012.1657 [nucl-ex].

\bibitem{v2ptb} D.~T.~Kettler, D.~J.~Prindle and T.~A.~Trainor,
  ``Transverse-rapidity $\bf y_t$ dependence of the nonjet azimuth quadrupole from 62 and 200 GeV Au-Au collisions,''
  Phys.\ Rev.\ C {\bf 91}, 064910 (2015),
arXiv:1504.02741 [hep-ph].

\bibitem{ppprd} J.~Adams {\it et al.}  (STAR Collaboration),
  ``The multiplicity dependence of inclusive p(t) spectra from p p  collisions
  Phys.\ Rev.\  D {\bf 74}, 032006 (2006),
  nucl-ex/0606028.

\bibitem{alicepbpb}  J.~Adam {\it et al.} (ALICE Collaboration),
``Centrality dependence of the nuclear modification factor of charged pions, kaons, and protons in Pb-Pb collisions at $\sqrt{s_{\rm NN}}=2.76$ TeV,''
Phys.\ Rev.\ C {\bf 93}, no. 3, 034913 (2016),
arXiv:1506.07287 [nucl-ex].



 
\end{thebibliography}
\end{document}